\documentclass[preprint2,times,tighten]{aastex6}
\pdfoutput=1 
\usepackage{amsmath,amstext}
\usepackage[figure,figure*]{hypcap}
\usepackage{newtxmath} 

\slugcomment{\it Draft --- 2017 February 16}

\def\deg{\ifmmode {^{\circ}}\else {$^\circ$}\fi}
\def\kms{\ifmmode {\rm\,km\,s^{-1}}\else
    ${\rm\,km\,s^{-1}}$\fi}
\def\ergcm2s{\ifmmode {\rm\,erg\,cm^{-2}\,s^{-1}}\else
    ${\rm\,erg\,cm^{-2}\,s^{-1}}$\fi}
\def\ergAcm2s{\ifmmode {\rm\,erg\,cm^{-2}\,s^{-1}\,\AA^{-1}}\else
    ${\rm\,erg\,cm^{-2}\,s^{-1}\,\AA^{-1}}$\fi}
\def\ergs{\ifmmode {\rm\,erg\,s^{-1}}\else
    ${\rm\,erg\,s^{-1}}$\fi}
\def\kmsMpc{\ifmmode {\rm\,km\,s^{-1}\,Mpc^{-1}}\else
    ${\rm\,km\,s^{-1}\,Mpc^{-1}}$\fi}

\def\oii{[\ion{O}{2}] $\lambda\lambda$3726,3729}
\def\oiis{[\ion{O}{2}]}

\def\oiiis{[\ion{O}{3}]}

\def\spose#1{\hbox to 0pt{#1\hss}}
\def\simlt{\mathrel{\spose{\lower 3pt\hbox{$\mathchar"218$}}
     \raise 2.0pt\hbox{$\mathchar"13C$}}}
\def\simgt{\mathrel{\spose{\lower 3pt\hbox{$\mathchar"218$}}
     \raise 2.0pt\hbox{$\mathchar"13E$}}}

\def\deg{$^{\rm o}$}

\def\arcsec{\ifmmode '' \else $''$\fi}

\def\arcsecpoint{\ifmmode ''\!. \else $''\!.$\fi}

\def\kms{\ifmmode {\rm km\ s}^{-1} \else km s$^{-1}$\fi}
\def\Msun{\ifmmode {\rm M}_{\odot} \else M$_{\odot}$\fi}
\def\Lsun{\ifmmode {\rm L}_{\odot} \else L$_{\odot}$\fi}
\def\Zsun{\ifmmode {\rm Z}_{\odot} \else Z$_{\odot}$\fi}

\def\ergscm2{ergs\,s$^{-1}$\,cm$^{-2}$}
\def\icm3{{\rm cm}^{-3}}
\def\icm2{{\rm cm}^{-2}}
\def\qo{\ifmmode q_{\rm o} \else $q_{\rm o}$\fi}
\def\Ho{\ifmmode H_{\rm o} \else $H_{\rm o}$\fi}
\def\ho{\ifmmode h_{\rm o} \else $h_{\rm o}$\fi}

\def\vFWHM{\ifmmode v_{\mbox{\tiny FWHM}} \else
            $v_{\mbox{\tiny FWHM}}$\fi}

\def\gtorder{\mathrel{\raise.3ex\hbox{$>$}\mkern-14mu
             \lower0.6ex\hbox{$\sim$}}}
\def\ltorder{\mathrel{\raise.3ex\hbox{$<$}\mkern-14mu
             \lower0.6ex\hbox{$\sim$}}}
\def\proptwid{\mathrel{\raise.3ex\hbox{$\propto$}\mkern-14mu
             \lower0.6ex\hbox{$\sim$}}}
\def\pz{P($z|C$)}

\begin{document}

\title{The Complete Calibration of the Color-Redshift Relation (C3R2) Survey:\\
Survey Overview and Data Release 1}

\author{Daniel C. Masters\altaffilmark{1}, Daniel K.  Stern\altaffilmark{2}, Judith G. Cohen\altaffilmark{3}, Peter L. Capak\altaffilmark{4},\\ Jason D. Rhodes\altaffilmark{2,7}, Francisco J. Castander\altaffilmark{5}, St\'ephane Paltani\altaffilmark{6}}


\altaffiltext{1}{Infrared Processing and Analysis Center, Pasadena,
CA 91125, USA}
\altaffiltext{2}{Jet Propulsion Laboratory, California Institute of Technology, Pasadena, CA 91109, USA}
\altaffiltext{3}{California Institute of Technology, Pasadena, CA 91125, USA}
\altaffiltext{4}{Spitzer Science Center, Pasadena, CA 91125, USA}
\altaffiltext{5}{Institut de Ci\`encies de l’Espai (ICE, IEEC/CSIC), E-08193 Bellaterra (Barcelona), Spain}
\altaffiltext{6}{Department of Astronomy, University of Geneva, Ch. d’Ecogia 16, 1290 Versoix, Switzerland}
\altaffiltext{7}{Kavli Institute for the Physics and Mathematics of the Universe, University of Tokyo, Chiba 277-8582, Japan}

\begin{abstract} 

A key goal of the Stage IV dark energy experiments  \emph{Euclid}, LSST  and \emph{WFIRST} is to measure the growth of structure with cosmic time from weak lensing analysis over large regions of the sky. Weak lensing cosmology
will be challenging: in addition to highly accurate galaxy shape
measurements, statistically robust and accurate photometric redshift (photo-z)
estimates for billions of faint galaxies will be needed in order to reconstruct the three-dimensional matter distribution. Here we present an overview of and initial results from the Complete Calibration of the Color-Redshift
Relation (C3R2) survey, designed specifically to calibrate the empirical galaxy
color-redshift relation to the \emph{Euclid} depth. These redshifts will also be important for the calibrations of LSST and \emph{WFIRST}. The C3R2 survey is obtaining multiplexed observations with Keck (DEIMOS,
LRIS, and MOSFIRE), the Gran Telescopio Canarias (GTC; OSIRIS), and the Very Large Telescope (VLT; FORS2 and KMOS) of a targeted sample of galaxies most important for the redshift calibration. We focus spectroscopic efforts on under-sampled regions of galaxy color space identified in previous work in order to minimize the number of spectroscopic redshifts needed to map the color-redshift relation to the required accuracy. Here we present the C3R2 survey strategy and initial results, including the 1283 high confidence redshifts obtained in the 2016A semester and released as Data Release 1. 

\end{abstract}


\section{Introduction}

The upcoming large-scale cosmology experiments \emph{Euclid} \citep{Laureijs11}, LSST \citep{Ivezic08} and \emph{WFIRST} \citep{Spergel15} will depend on robust photometric redshift (photo-z) estimates for billions of faint galaxies in order to obtain a three-dimensional picture of the growth of cosmic structure. Small ($\gtrsim$0.2\%) redshift biases can easily dominate the overall error budget in the cosmological parameters measured by these surveys (e.g., \citealp{Huterer06}). Spectroscopic calibration efforts for these missions must therefore measure the color-redshift relation of galaxies with sufficient fidelity to reconstruct the redshift distributions of shear samples with negligible systematic bias. While photometric redshift estimation techniques have grown in sophistication and precision over the past few decades (e.g., \citealp{Benitez00, Brammer08, Ilbert09,Carrasco13, Speagle16}), existing methods have not met the photo-z accuracy requirements set by weak lensing cosmology. 

The relation of seven or eight galaxy broadband colors (referred to henceforth by the vector $C$) to redshift is ultimately an empirical question. In \citet{Masters15} (hereafter M15) we demonstrated a method, based on the \emph{self-organizing map} (SOM; \citealp{Kohonen90}) algorithm, to constrain the empirical multidimensional color distribution of galaxies present in a survey. This technique allowed us to project the multicolor distribution of galaxies in a topologically ordered way onto a two-dimensional grid. By applying this technique to a well-studied deep field with uniform \emph{ugrizYJH} photometry, we were able to demonstrate that spectroscopic surveys to date do not sample the full color-space of galaxies in a \emph{Euclid}-like survey, and thus the color-redshift relation is not fully constrained with existing spectroscopy. This issue is of particular concern for machine learning-based photo-z estimation, which requires color-complete training samples, but also affects the calibration of template-based techniques. 

The analysis in M15 motivated a survey designed to systematically map the color-redshift relation over the currently undersampled regions of galaxy color space relevant to \emph{Euclid}. M15 estimated that $\sim$5000 new redshifts, carefully distributed in color space, would be sufficient to meet the stringent requirements for weak lensing cosmology. This ``direct" approach to photo-z calibration is complementary to approaches based on spatial cross-correlation of photometric samples with spectroscopic samples (e.g., \citealt{Newman08,Rahman15}. At least two independent methods to measure $N(z)$ for the tomographic shear samples will be required to ensure no systematic photo-z biases exist; these methods can therefore serve as useful checks on each other. 

Here we describe the initial stage of what we are calling the Complete Calibration of the Color-Redshift Relation (C3R2) survey, designed to fill out the color space of galaxies with secure redshifts to the \emph{Euclid} weak lensing depth. By doing so, the empirical \pz\ relation obeyed by galaxies can be constrained with sufficient accuracy to meet the cosmological requirements of \emph{Euclid}. The spectra will also be of significant value for the LSST and \emph{WFIRST} calibrations, which will be more difficult than for \emph{Euclid} due to the greater photometric depth of those surveys (Hemmati et al, 2017, in prep). We estimate that $\sim$40 Keck nights in total (or their equivalent) could achieve the fidelity required to meet the cosmological requirements for \emph{Euclid}, when combined with extensive existing spectroscopy.

\begin{figure*}[htb]
\centering
  \begin{tabular}{@{}ccc@{}}
    \includegraphics[clip, trim=1cm .5cm 1cm 0.5cm, width=.32\textwidth]{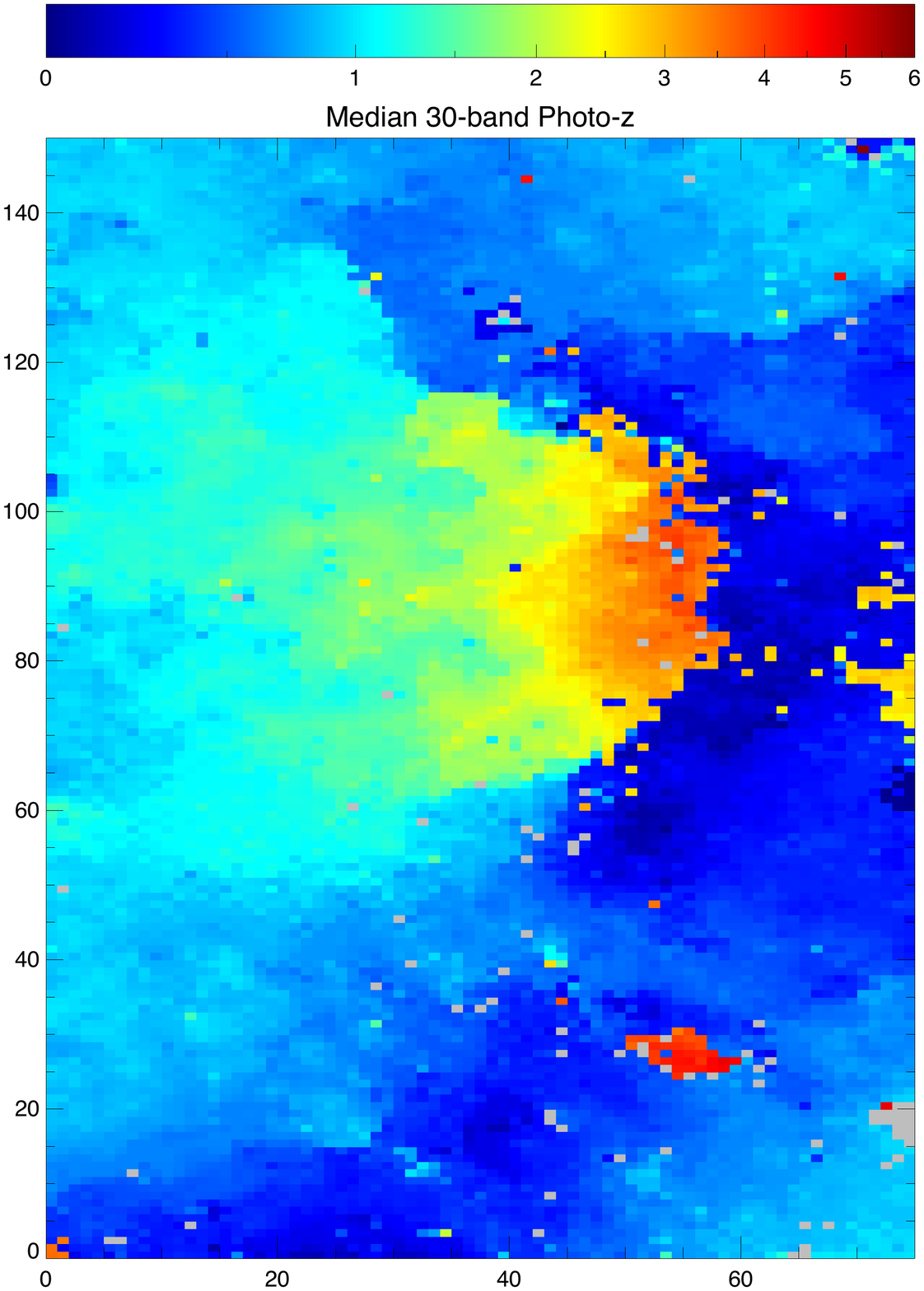} &
    \includegraphics[clip, trim=1cm .5cm 1cm 0.5cm, width=.32\textwidth]{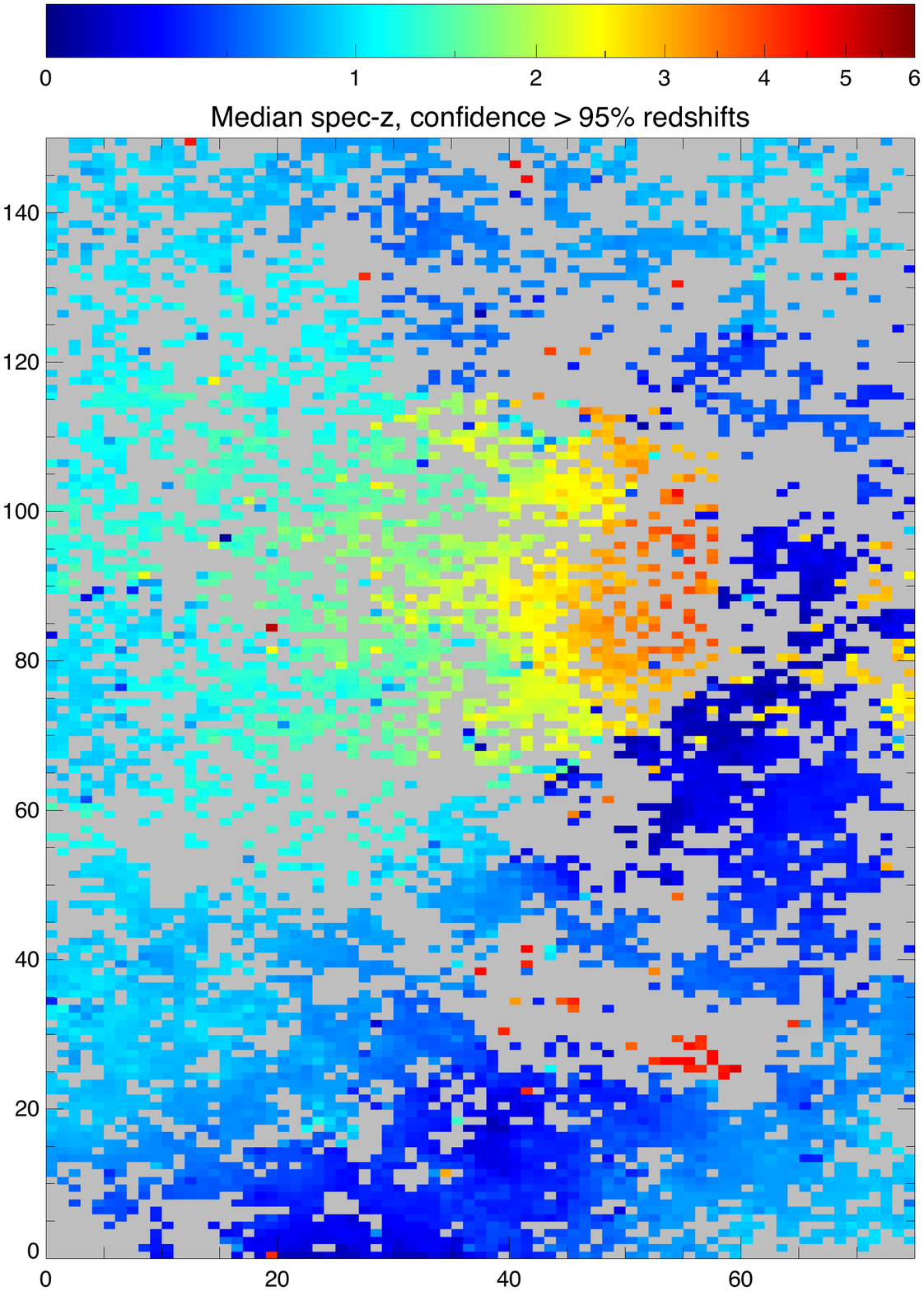} &
    \includegraphics[clip, trim=1cm .5cm 1cm 0.5cm, width=.32\textwidth]{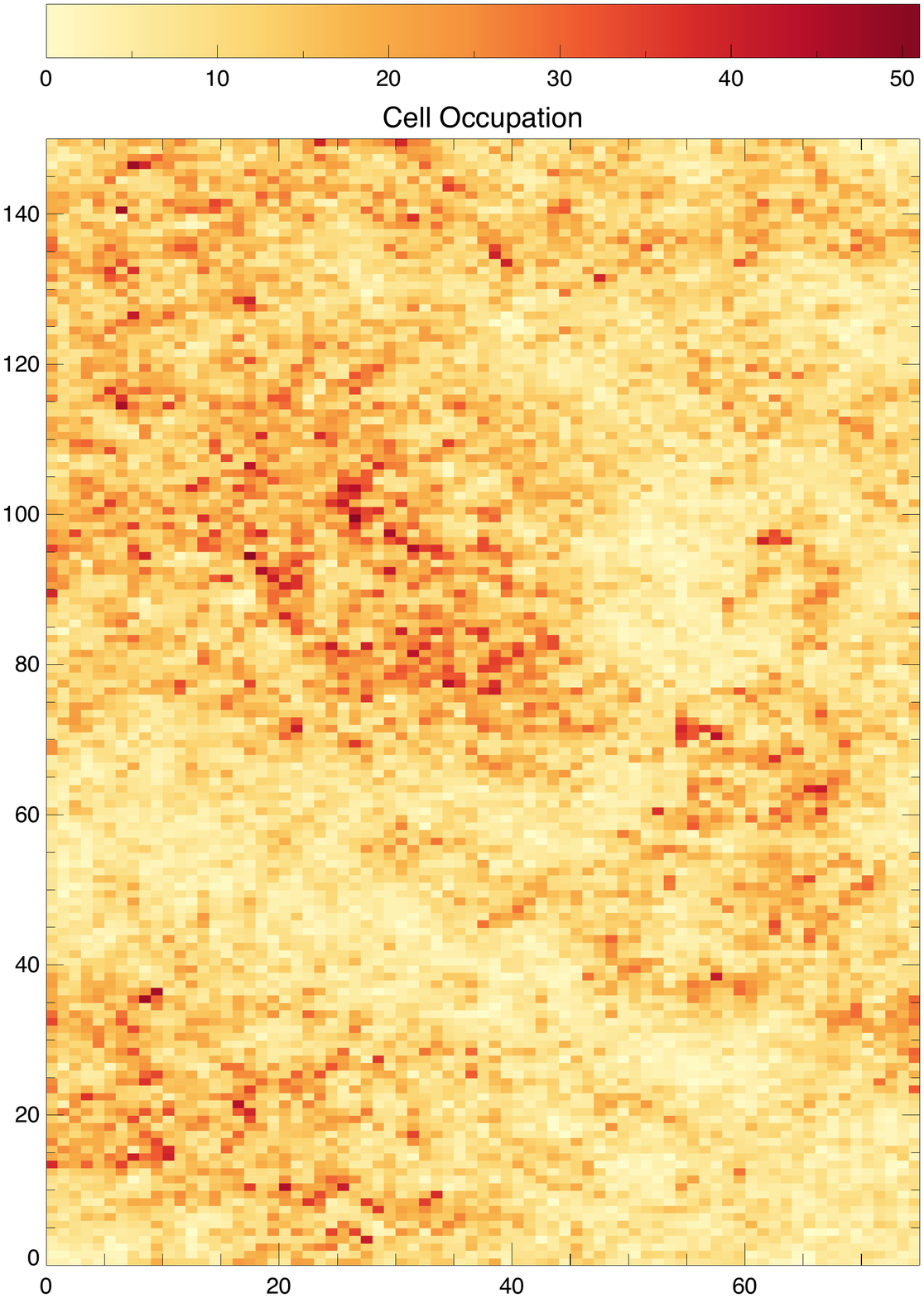} \\
  \end{tabular}
  \label{figure:colors}
  \caption{The self-organizing map (SOM) forming the basis for the C3R2 targeting strategy in the 2016A semester. Each one 	of the 11,250 cells represents a particular galaxy SED that shows up in the data with some degree of regularity; see \citet{Masters15} for details. Note that the axes are not physical, but are merely indices into the map. {\bf \emph{Left:}} The map colored by the median 30-band photometric redshifts of galaxies associating with each cell, using the full input sample of sources. {\bf \emph{Center:}} Our imperfect knowledge of the \emph{empirical} \pz\ mapping from existing spectroscopy in the field. The cells are colored according to the median redshift of spec-z objects in the color cell with high-confidence redshifts. Gray cells have no galaxies with existing high-confidence redshifts. {\bf \emph{Right:}} The density of galaxies across the map. These are the key ingredients of the C3R2 survey strategy.}
\end{figure*}

This paper gives an overview of the C3R2 survey and presents results from the 2016A semester, which constituted the first five nights of observing. All 2016A observations were done with Keck. The structure of this paper is as follows. In \S2 we give an overview of C3R2 survey strategy. In \S3 we discuss the observations and data reduction for the first five nights of observing. In \S4 we describe redshift determination and the identification of serendipitous sources. In \S5 we present initial results from the survey. In \S6 we conclude with a discussion. High confidence redshifts from DR1 are provided in a machine readable table.

\section{C3R2 Survey Overview}

The Keck portion of the C3R2 survey is a joint effort between Caltech (PI: J. Cohen), NASA (PI: D. Stern), the University of California (PI: B. Mobasher), the University of Hawaii (PI: D. Sanders). European participation in C3R2 with the GTC (PI: C. Gutierrez) and VLT (PI: F. Castander), as well as Harvard participation with MMT (PI: D. Eisenstein), will commence in 2017. The first five nights of observing with Keck were allocated by Caltech in the 2016A semester. Here we provide a brief overview of the C3R2 strategy for these observations. 

\subsection{The Self-Organized Map of Galaxy Colors}

In M15 we used COSMOS \citep{Capak07, Scoville07, Laigle16} \emph{ugrizYJH} photometry of $\sim$130k galaxies, closely resembling what will be obtained by the \emph{Euclid} survey, to map the color distribution of galaxies to the \emph{Euclid} depth ($i$$\sim$24.5~AB). We used the SOM algorithm (a manifold learning technique for nonlinear dimensionality reduction) to generate a topologically ordered 2D representation of the high-dimensional color distribution\footnote{The SOM algorithm was used mainly for its relative simplicity and visualization power; however, any technique that manages to quantify the density of galaxies in multicolor space would be equally appropriate. A number of other techniques for nonlinear dimensionality reduction (e.g., the generative topographic map, growing neural gas, and local linear embedding) may, in principle, offer some advantages over the SOM.}.  Galaxies from COSMOS were then matched back to the self-organized map according to their best-matching color cell in the SOM. This sorting of galaxies enables a variety of analyses, including the density of galaxies in different parts of color space, the median 30-band photometric redshifts from COSMOS as a function of position in color space, and the distribution of spectroscopic redshifts on the map (Figure~\ref{figure:colors}). Importantly, by placing all existing spectroscopy from the COSMOS field on the map, we reveal regions of color space for which no galaxies have existing high-confidence redshifts. Of greatest importance for the C3R2 survey are: (1) the current spectroscopic sampling across color space, and (2) the source density as a function of position in color space, as more common galaxies will contribute more to the cosmic shear signal.

\subsection{Existing Spectroscopy Across Galaxy Color Space}

For C3R2 we need to identify the regions of galaxy color space for which spectroscopic redshifts already exist and where they are systematically missing. We collected existing spectroscopy in COSMOS to do this, as described in M15. These redshifts include (but are not limited to) those from VLT-VIMOS \citep{Lilly07,LeFevre15}, Keck-MOSFIRE \citep{Kriek15}, Keck-DEIMOS \citep{Kartaltepe10}, and Magellan-IMACS \citep{Trump07}. For the 2016A run we used only the spectroscopy taken in the COSMOS survey to identify undersampled regions of color space. The reason we could not incorporate spectroscopy from other fields for these observations is that the photometry between fields has to be highly consistent in multiple bands to reliably place galaxies on the same color map; at the time this problem had not been solved. Significant subsequent work has been done to solve this problem for upcoming runs, to be described in a forthcoming paper. The fields that have subsequently been put on a highly consistent color frame in \emph{ugrizYJH} to the \emph{Euclid} depth are VVDS, SXDS, and EGS (in addition to COSMOS). 

\subsection{Target Prioritization}

For the 2016A observations we used the SOM derived in M15 to prioritize regions of galaxy multicolor space that are currently undersampled by existing spectroscopic surveys. For observed fields in 2016A other than COSMOS (SXDS and EGS), we attempted to bring the photometry on to the COSMOS color system in order to select the targets in a consistent way. We used the CANDELS \citep{Grogin11} photometry in a subset of the COSMOS field together with the CANDELS photometry in SXDS and EGS to derive a rough color conversion between the fields.

Target prioritization for C3R2 is based on two main factors: (1) the usefulness of a galaxy for calibrating the \pz\ relation, and (2) the likelihood of obtaining a secure redshift given the instrument, exposure time, and expected galaxy properties. The usefulness of a particular galaxy to the redshift calibration effort depends both on how common its colors are in the data and whether high-confidence redshifts for galaxies with similar colors already exist. 

Based on these considerations, we developed a prioritization scheme for galaxies that weights sources in unsampled cells of the SOM more heavily, and also gives preference to more common galaxy colors. The priorities for C3R2 are adaptive as new data is obtained and more of the color space is filled in. For the 2016A run our priority scheme was as follows:
\begin{enumerate}
 \item{We assign a initial priority value of 10 to objects occupying cells with no spectroscopic redshifts of even moderate quality (the gray regions of the SOM in the middle panel of Figure~\ref{figure:colors}), a starting priority of 3 to objects in cells with a spectroscopic redshift(s) of only moderate confidence, and a starting priority of 1 to galaxies in color cells that already have one or more high-confidence spectroscopic redshifts. Galaxies with existing redshifts of at least moderate confidence were not targeted.} \item{We multiply each galaxy's priority by the number of objects in its color cell, effectively upweighting sources with common SEDs.\footnote{We have since substantially lessened the extent to which we weight by cell occupation, because it is effectively accomplished already by the source density on the sky - i.e., more common sources will find their way onto masks more frequently.}} \item{We penalize objects that are color outliers within their color cell in order to avoid using them for calibration. A small fraction of objects in the sample are not represented well in the SOM, either because they have abnormal colors from photometric errors or the superposition of two or more  sources, or are truly rare objects (e.g., X-ray sources). We want to avoid calibrating with these.}
\end{enumerate}

As will be described in future data releases, this prioritization scheme has been refined for the 2016B and later observations to more efficiently map the color-redshift relation. In M15 we pointed out that spectroscopic effort could also be intentionally directed at regions of color space with intrinsically higher redshift uncertainty (e.g., with double-peaked redshift PDFs). For now we have not prioritized based on redshift uncertainty; however, as the survey progresses and the color map is filled in we may incorporate this quantity. 

\subsection{Estimating Required Instruments \& Exposure Times}

A crucial element of the C3R2 survey is the use of best-fit spectral templates to the galaxies to predict the exposure times with different instruments needed to obtain a secure redshift. If we then fail to obtain a redshift under nominal observing conditions we can prioritize the target further for follow-up. This potential re-targeting is important to avoid systematic biases in the redshifts obtained in different parts of color space.

We use a technique developed for the proposed SPHEREx mission \citep{Dore14,Stickley16} to predict the spectrum of galaxies based on their broadband photometry. In brief, this method fits a set of templates based on the libraries of \citet{Brown14} (for galaxies) and \citet{Salvato09} (for AGN) to deep multiband photometry. Based on the analysis in \citet{Stickley16}, we can estimate the continuum to within 20\% and the emission line strengths to within a factor of two.  We then use the instrumental response curves for each telescope and instrument to estimate the required integration time to obtain a redshift to that galaxy, given its estimated photometric redshift. Primary objects for a mask are those expected to yield a redshift within a factor of two of the intended mask integration time. The time estimates were compared with previous observations to verify their accuracy. As described in \S4.2, we use a flagging scheme to keep track of objects for which a redshift was expected but not obtained. These sources can then be prioritized for additional observations.

\begin{deluxetable*}{lllcl}
\tablewidth{0pt}
\tablecaption{List of observing nights.}
\tablehead{
\colhead{UT Date} &
\colhead{Code} &
\colhead{Instrument} &
\colhead{\# Masks} &
\colhead{Observing conditions}}
\startdata
2015 Dec 15 & N01-D & DEIMOS  & 4 & clear, 0\farcs65 seeing \\
2016 Feb 28 & N02-M & MOSFIRE & 6 & clear, 0\farcs5-0\farcs65 seeing \\
2016 Feb 29 & N03-D & DEIMOS  & 4 & clear, 0\farcs65 seeing; moon \\
2016 Mar 01 & N04-D & DEIMOS  & 7 & clear, 1\farcs0 seeing; moon \\
2016 Apr 09 & N05-L & LRIS    & 4 & thin cirrus, 0\farcs97 seeing 
\enddata
\label{table:obsevations}
\end{deluxetable*}

\section{Observations and Data Reductions}

Five nights were allocated by Caltech in the 2016A semester: three nights on DEIMOS \citep{Faber03}, and one night each with LRIS \citep{Oke95} and MOSFIRE \citep{McLean12}. Tables 1 and 2 summarize the nights and observed slitmasks. All five nights had excellent observing conditions. Here we describe the observations and data reduction.

\subsection{DEIMOS}

DEIMOS observations were conducted using the 600~groove~mm$^{-1}$ grating blazed at 7200~\AA\ and the GG400 blocking filter, with dithering performed to improve  sky subtraction. In the initial observing run, we experimented with minimum slit lengths of both 6$''$ and 10$''$, with no significant difference in redshift success rate. In the subsequent DEIMOS observations we settled on a minimum slit width of 8$''$ as a balance between getting the most targets possible on the mask and getting good sky measurements. Data were reduced using a modified version of the DEEP2 pipeline  designed to deal with dithered data. 

\subsection{LRIS}

We used the 400 groove~mm$^{-1}$ blue grism blazed at 3400~\AA\ and the 400 groove~mm$^{-1}$ red grating blazed at 8500~\AA, with the D560 dichroic. Our choice of blue grism gives high sensitivity at bluer wavelengths where identifying features are likely to be found for objects with photometric redshifts of $z\sim1.5-3$, while the red coverage allows for the detection of \oiis\ for some sources out to $z\sim1.6$. The LRIS spectra were reduced using the IRAF-based BOGUS software developed by D. Stern, S. A. Stanford, and A. Bunker, and flux calibrated using observations of standard stars from  \citet{Massey90} observed on the same night using the same instrument configuration.

\subsection{MOSFIRE}

MOSFIRE was used in its default configuration. For instrumental details we refer the reader to \citet{Steidel14}. We observed four masks in $Y$ band and two in $K$ band, using integration times of 180s with ABAB dithering to improve sky subtraction. Reductions were performed with the MOSFIRE Data Reduction Pipeline (DRP) made available by the instrument team\footnote{https://keck-datareductionpipelines.github.io/MosfireDRP/}.

\section{Redshift Determination}

Each observed source was assessed independently by two co-authors to determine the redshift and associated quality flag. These results were then compared for conflicts in either redshift or quality flag. Conflicts were reconciled through a joint review of the spectra, usually with the help of a third, independent reviewer.  As a final step in the process, we investigated all Q=4 (highest quality, see \S4.1) sources for which the spectroscopic redshift ($z_s$) was highly discrepant from the expected photometric redshift ($z_p$, defined as the median photometric redshift of sources in the relevant SOM cell).  Specifically, we investigated all sources with $\lvert z_{p} - z_{s} \rvert / (1 + z_{s}) \geq 0.15$.  For most of these outliers, the spectroscopic redshift was deemed solid and we discuss the nature of the discrepancy in more detail in \S~5.2.  However, for two cases, this step caused us to modify the final redshift assessment.  One of these final modifications was due to confusing a target and very close ($\sim 1\arcsec$ separation) serendipitous source, while the other modification was due to a genuine error in line identification aggravated by incomplete sky-line subtraction mimicking a corroborating emission line.

\begin{figure*}[htb]
\centering
 \includegraphics[clip, trim=1.5cm .5cm .5cm 1cm, width=0.79\textwidth]{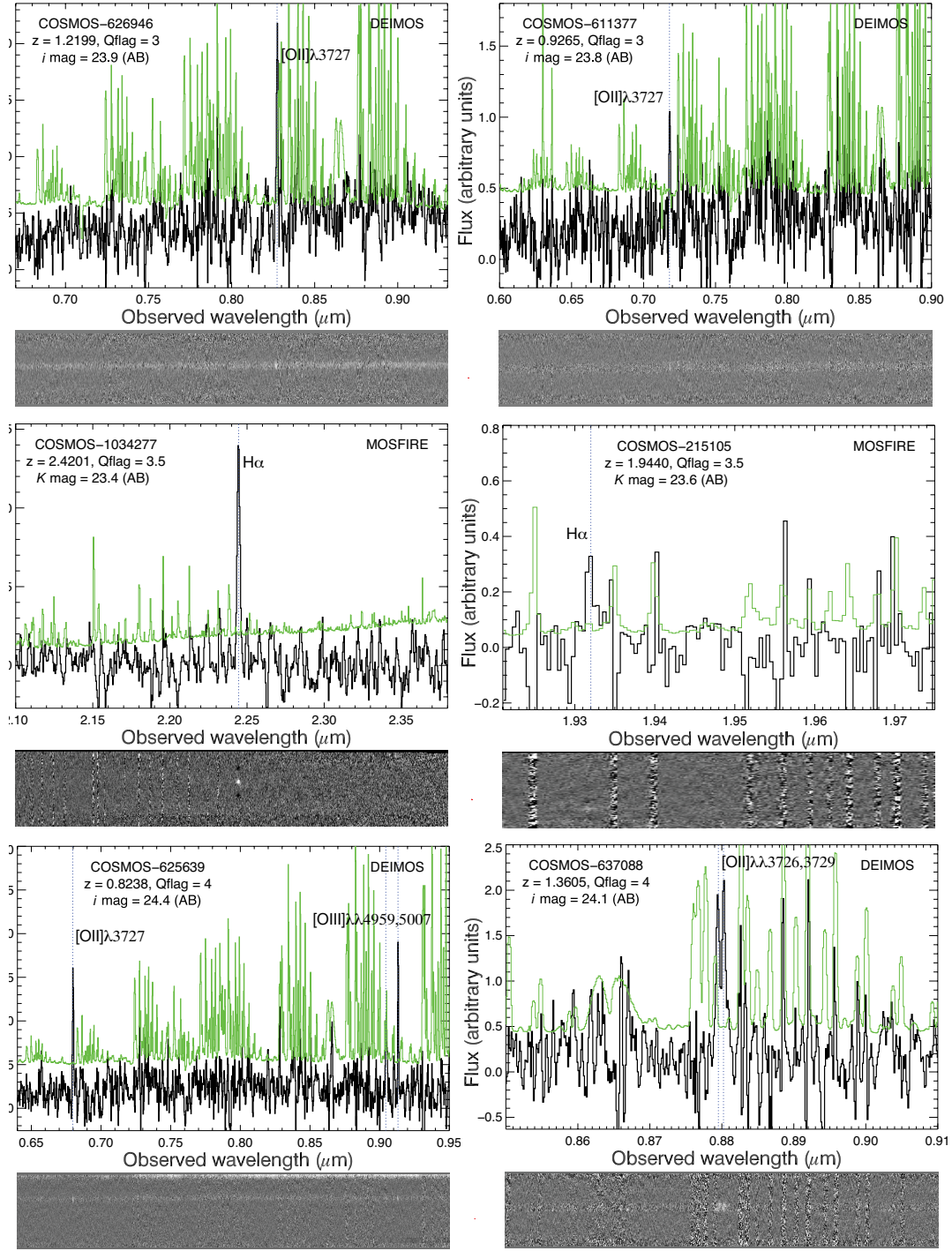} 
  \caption{Example spectra from the 2016A semester. Error spectra are overplotted in green. {\bf \emph{Top, left:}} DEIMOS spectrum with a confidence flag Q=3 redshift based on a line identified as \oiis\ with an associated continuum break. The redshift is not class Q=4 as the line is not well-resolved into the \oii\ doublet. {\bf \emph{Top, right:}} Another similar example of a Q=3 redshift based on a weak line identified as \oiis\ together with a continuum break.   
{\bf \emph{Middle, left:}}  A MOSFIRE $K$-band spectrum of a source at z=2.4201. The single strong line is identified as H$\alpha$, with a confidence flag Q=3.5 indicating high confidence (but not certainty, due to the lack of corroborating features).  {\bf \emph{Middle, right:}} A similar example of a Q=3.5 MOSFIRE redshift based on a weak line identified as H$\alpha$. {\bf \emph{Bottom, left:}} DEIMOS spectrum illustrating the assignment of a confidence flag Q=4 redshift based on the detection of multiple features (\oiis$\lambda$3727, \oiiis$\lambda\lambda$4959,5007).  {\bf \emph{Bottom, right:}} DEIMOS spectrum illustrating a Q=4 redshift based on the clear, resolved presence of the \oii\  doublet.}
      \label{figure:example_spectra}
\end{figure*}

\subsection{Quality Flags
}
The redshift flagging scheme we use is similar to that adopted by the zCOSMOS \citep{Lilly07}, DEEP2 \citep{Newman13}, and VUDS surveys \citep{LeFevre15}. The quality flags range from 0-4 with 4 indicating the highest confidence redshift and 0 indicating that no redshift could be found. The interpretation of the flags is roughly as follows: 

\scriptsize
\begin{deluxetable*}{lcccccc}
\tablecaption{List of observed slitmasks.}
\tablehead{
\colhead{} &
\colhead{} &
\colhead{R.A.} &
\colhead{Dec.} &
\colhead{PA} &
\colhead{Exposure} &
\colhead{\# Targets} \\
\colhead{Mask ID / Name} &
\colhead{Night} &
\colhead{(J2000)} &
\colhead{(J2000)} &
\colhead{(\deg)} &
\colhead{(s)} &
\colhead{(total / Q=4 / ser)}}
\startdata
16A-D01 / UDS-m1n1    & N01-D &  2:17:27.0 &  $-$5:15:07 &     90.0 & 2$\times$1800 &  86 / 51 /  5 \\
16A-D02 / UDS-m3n1    & N01-D &  2:17:27.0 &  $-$5:14:07 &     90.0 & 2$\times$1800 & 100 / 66 /  5 \\
16A-D03 / COSMOS-m3n1 & N01-D & 10:00:22.0 &  $+$2:20:00 &     90.0 & 4$\times$1800 & 104 / 72 /  7 \\
16A-D04 / COSMOS-m4n1 & N01-D & 10:00:22.0 &  $+$2:35:00 &     90.0 & 4$\times$1800 &  70 / 60 / 19 \\
16A-M05 / COSMOS-m1-Y & N02-M & 10:00:57.2 &  $+$1:48:40 &     85.0 & 20$\times$180 &  24 / 8 / 1 \\
16A-M06 / COSMOS-m2-Y & N02-M & 10:00:54.4 &  $+$2:01:47 &     55.0 & 20$\times$180 &  29 / 3 / 0 \\
16A-M07 / COSMOS-m3-Y & N02-M & 10:00:57.7 &  $+$2:14:38 &     40.0 & 20$\times$180 & 24 / 5 / 1 \\
16A-M08 / COSMOS-m1-K & N02-M & 10:00:10.5 &  $+$2:14:20 &     30.0 & 20$\times$180 & 12 / 5 /  0 \\
16A-M09 / COSMOS-m4-Y & N02-M & 10:00:14.2 &  $+$2:03:34 &     30.0 & 16$\times$180 & 25 / 6 / 1 \\
16A-M10 / EGS-m1-K    & N02-M & 14:17:57.4 & $+$52:35:51 &     25.0 & 22$\times$180 & 23 / 3 / 0 \\
16A-D11 / COSMOS-m1n2 & N03-D &  9:58:43.2 &  $+$1:42:00 &     90.0 & 3$\times$1200 & 93 / 62 / 2 \\
16A-D12 / COSMOS-m8n2 & N03-D &  9:58:43.3 &  $+$2:12:47 &     90.0 & 6$\times$1200 & 92 / 56 / 17 \\
16A-D13 / COSMOS-m2n2 & N03-D &  9:58:43.2 &  $+$1:46:15 &     90.0 & 3$\times$1200 & 91 / 77 / 7 \\
16A-D14 / COSMOS-m9n2 & N03-D &  9:58:43.2 &  $+$2:17:00 &     90.0 & 6$\times$1200 & 99 / 64 / 3 \\
16A-D16 / COSMOS-m3n2 & N04-D &  9:58:43.2 &  $+$1:50:24 &     90.0 & 3$\times$1200 & 89 / 43 / 4 \\
16A-D17 / COSMOS-m4n2 & N04-D &  9:58:43.2 &  $+$1:54:36 &     90.0 & 3$\times$1200 & 95 / 70 / 4 \\
16A-D18 / COSMOS-m7n2 & N04-D &  9:58:43.2 &  $+$2:08:15 &     90.0 & 6$\times$1200 & 91 / 52 / 13 \\
16A-D19 / COSMOS-m6n2 & N04-D &  9:58:43.2 &  $+$2:04:16 &     90.0 & 4$\times$1200 & 94 / 72 / 9 \\
16A-D20 / COSMOS-m5n2 & N04-D &  9:58:43.2 &  $+$1:58:48 &     90.0 & 3$\times$1200 & 98 / 80 / 3 \\
16A-D21 / EGS-m1n2    & N04-D & 14:18:00.0 & $+$52:33:00 &     90.0 & 3$\times$1200 & 100 / 62 / 10 \\
16A-D22 / EGS-m2n2    & N04-D & 14:18:00.0 & $+$52:41:24 &     90.0 & 3$\times$1200 & 104 / 72 / 3 \\
16A-L23 / COSMOS-m1n5 & N05-L &  9:59:44.1 &  $+$2:36:12 &  $-$60.0 & 4$\times$1200 & 25 / 3 / 1 \\
16A-L24 / COSMOS-m3n5 & N05-L &  9:58:58.7 &  $+$2:45:56 & $-$110.0 & 2$\times$1200 & 18 / 8 / 0 \\
16A-L25 / EGS-m1n5    & N05-L & 14:19:08.6 & $+$52:28:48 &      0.0 & 5$\times$1200 & 28 / 11 / 0 \\
16A-L26 / EGS-m2n5    & N05-L & 14:18:04.8 & $+$52:42:01 &      0.0 & 5$\times$1200 & 26 / 4 / 1  
\enddata
\label{table:slitmasks}
\tablecomments{`Night' column refers to observing code in second
column of Table~1: night number, followed by letter indicating
instrument used (D -- DEIMOS, L -- LRIS, M -- MOSFIRE).  R.A. and
Dec.  refer to the mask center.  Final column gives total number
of slitlets in mask (ignoring those with failure code $-93$/$-94$ as described in \S4.2), total number of high-quality (Q=4) redshifts
measured, and the number of serendipitous sources with high-quality
redshifts (quality flag Q = 4).}
\end{deluxetable*}
\normalsize 
\begin{itemize}
\item{Q=4: A quality flag of 4 indicates an unambiguous redshift identified with multiple features or the presence of the split \oii\ doublet.}
\item{Q=3.5: A quality flag of 3.5 indicates a high-confidence redshift based on a single line, with a very remote possibility of an incorrect identification. An example might be a strong, isolated emission line identified as H$\alpha$, where other identifications of the line are highly improbable due to the lack of associated lines or continuum breaks. This flag is typically only adopted for LRIS and MOSFIRE spectra.}
\item{Q=3: A quality flag of 3 indicates a high-confidence redshift with a low probability of an incorrect identification. An example might be the low signal-to-noise ratio detection of an emission line, possibly corrupted by telluric emission or absorption, identified as \oii, but where the data quality is insufficient to clearly resolve the doublet.}
\item{Q=2/1: A quality flag of 2 indicates a reasonable guess, while a quality flag of 1 indicates a highly uncertain guess.  Sources with these low confidence redshifts are not included in the data release.}
\item{Q=0: A quality flag of 0 indicates that no redshift could be identified. As described next, a code indicating the cause of the redshift failure is assigned in place of the redshift.}
\end{itemize}

\begin{figure*}[htb]
\centering
  \begin{tabular}{@{}cc@{}}
 \includegraphics[clip, trim=1cm 5cm 0.5cm 6cm, width=0.49\textwidth]{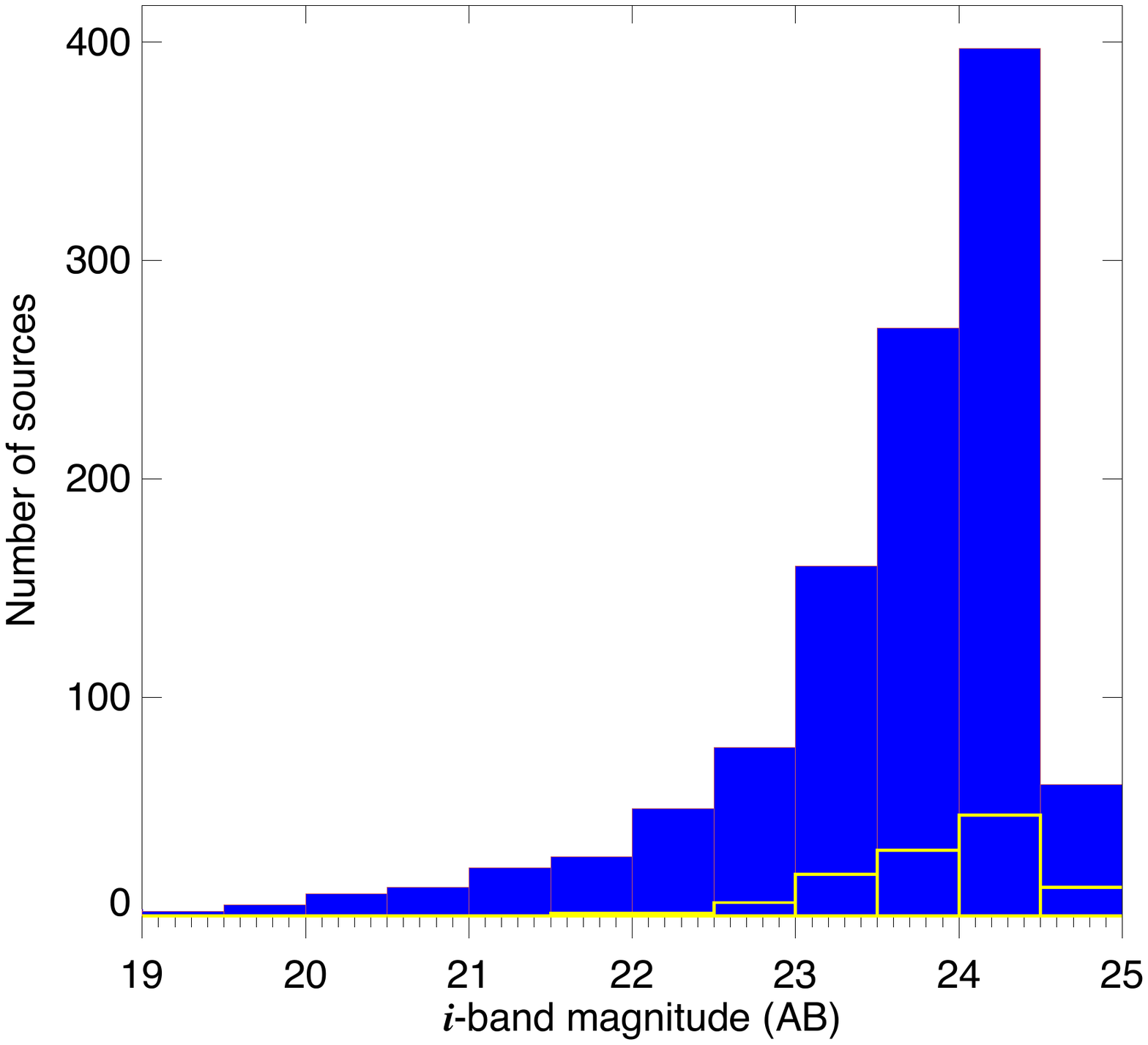} &
    \includegraphics[clip, trim=1cm 5cm 0.5cm 6cm, width=.49\textwidth]{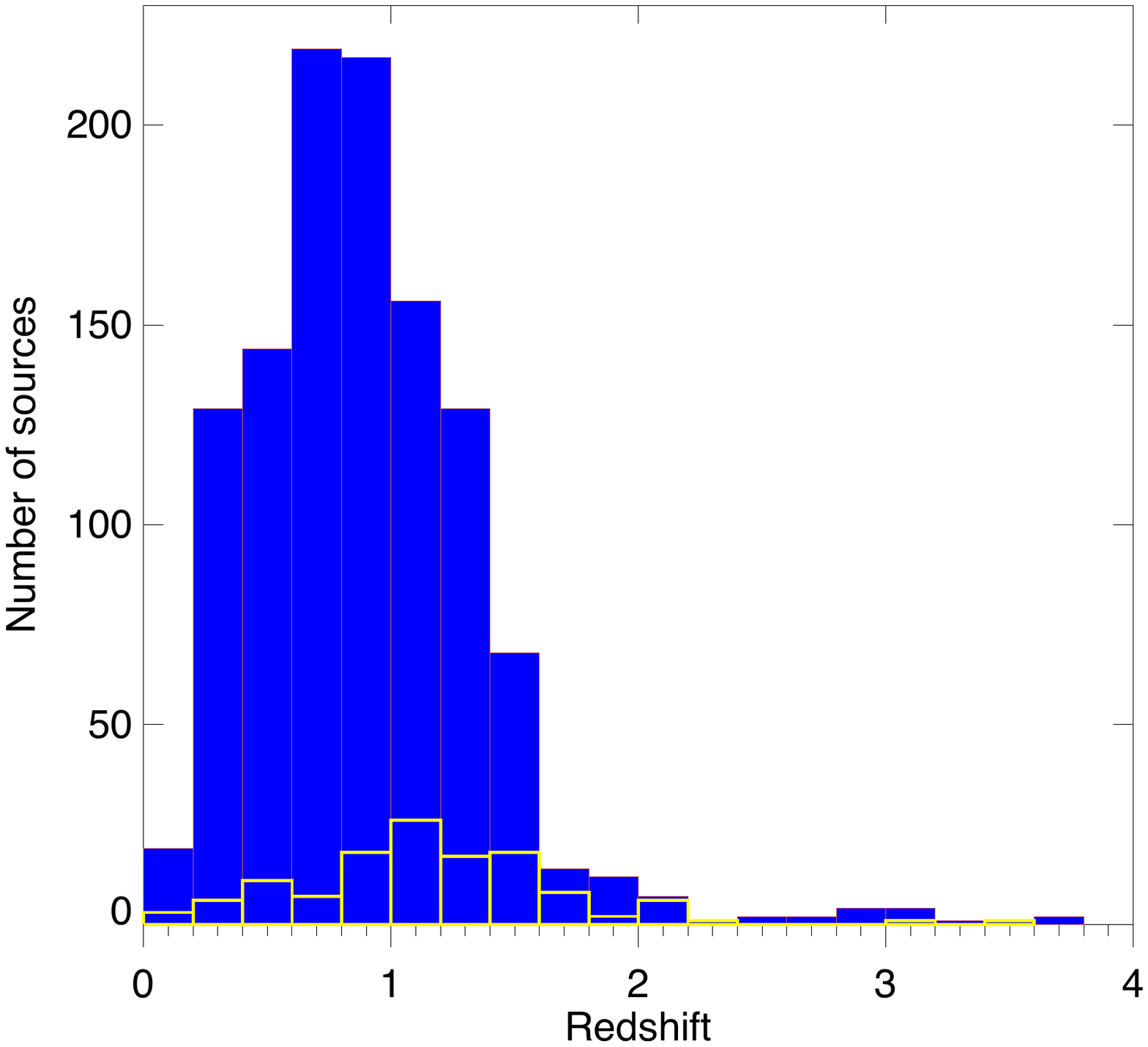} \\
  \end{tabular}
  \caption{{\bf \emph{Left:}} The \emph{i}-band magnitude distribution of C3R2-targeted sources yielding Q=4 redshifts (blue) and Q=3/3.5 redshifts (yellow) in the 2016A semester (1,283 sources total). The majority are fainter than \emph{i} = 23 (AB). {\bf \emph{Right:}} The redshift distribution of the same samples.}
      \label{figure:mag_and_z_dist}
\end{figure*}

Figure~\ref{figure:example_spectra} shows C3R2 spectra from 2016A as examples of Q=3, Q=3.5, and Q=4 redshift assignments.


\subsection{Failure Codes}


It is important for C3R2 to track redshift failures, as well as the reasons for the failures, in order to avoid systematic biases in the sources selected for calibration. Failed targets that were expected to yield a redshift given the instrument and exposure time can be prioritized for additional follow-up. On the other hand, if no spectroscopic redshift was obtained because of a problem with the observing conditions or data (i.e., bad rows, or the target ended up in a region between two detector arrays), no additional prioritization of that source may be needed. 

With these considerations in mind, we developed a system to flag different ``failure modes'' for objects not yielding a redshift. Four categories of failures are used, with the corresponding code assigned in place of a redshift in our catalog. The failure modes we identify are:

\begin{enumerate}
\setlength\itemsep{0.5em}
\item{Code = $-$91: Object too faint to identify the redshift. Indicates that a deeper exposure and/or different instrument and/or different wavelength coverage is required to obtain a secure redshift. An example of such a source might be a galaxy expected to have a strong \oiis\ emission line at 9800 \AA, but where the slit placement caused the wavelength coverage to end at 9500 \AA, yielding a continuum detection without any strong spectroscopic features.  These sources will be further prioritized in future observations.}
\item{Code = $-$92: Object well-detected, but no redshift could be determined. May require a different instrument for secure redshift determination due to an incorrect photometric redshift or the wavelength coverage obtained for a given observation.  We emphasize that the dividing line between $-$91 and $-$92 failure codes is imprecise, and no strong effort was made to homogenize the classification.  Fundamentally, both codes can be considered as two aspects of the same issue.  Again, these sources will increase in priority going forward.}
\item{Code = $-$93: Corrupted slit, typically due to bad rows/columns in the data or the source falling on or near detector chip gaps. Does not affect object priority in future observations.} 
\item{Code = $-$94: Missing slit, as an extreme case of code $-$93. Does not affect object priority in future observations.}
\end{enumerate}

\noindent{Failure codes $-$91 and $-$92 essentially correspond to spectral quality issues (i.e., signal-to-noise ratio, wavelength range), while codes $-$93 and $-$94 correspond to data quality issues (i.e., slitmask design issues, detector issues).  While in DR1 we distinguished between these four failure modes, for many analysis purposes, considering just the two general categories will be sufficient.}

Failure code $-$91, the most common failure code, generally indicates that the signal-to-noise ratio of the data was insufficient for redshift determination. Indeed, considering the 131 DEIMOS-observed sources in COSMOS with this failure code, 122 (93\%) were \emph{anticipated} to fail based on our estimated exposure time needed to get a redshift. As with low-confidence redshifts, sources for which we failed to find a redshift are not included in this data release.

\begin{figure}[htb]
\centering
  \begin{tabular}{@{}c@{}}
    \includegraphics[clip, trim=0.5cm 6.3cm 0.5cm 6cm, width=.5\textwidth]{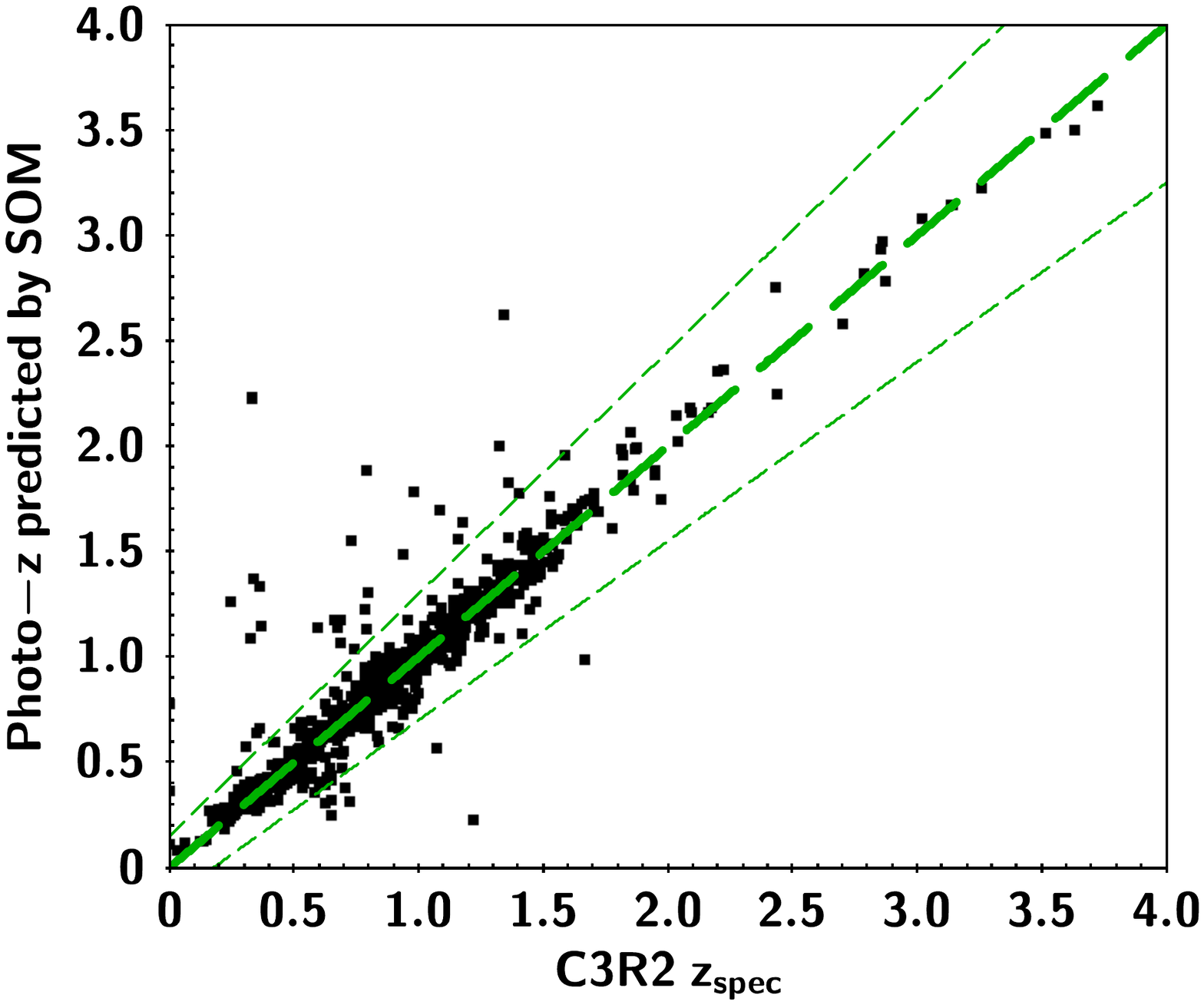} \\
    \includegraphics[clip, trim=0.5cm 6.3cm 1cm 6cm, width=.5\textwidth]{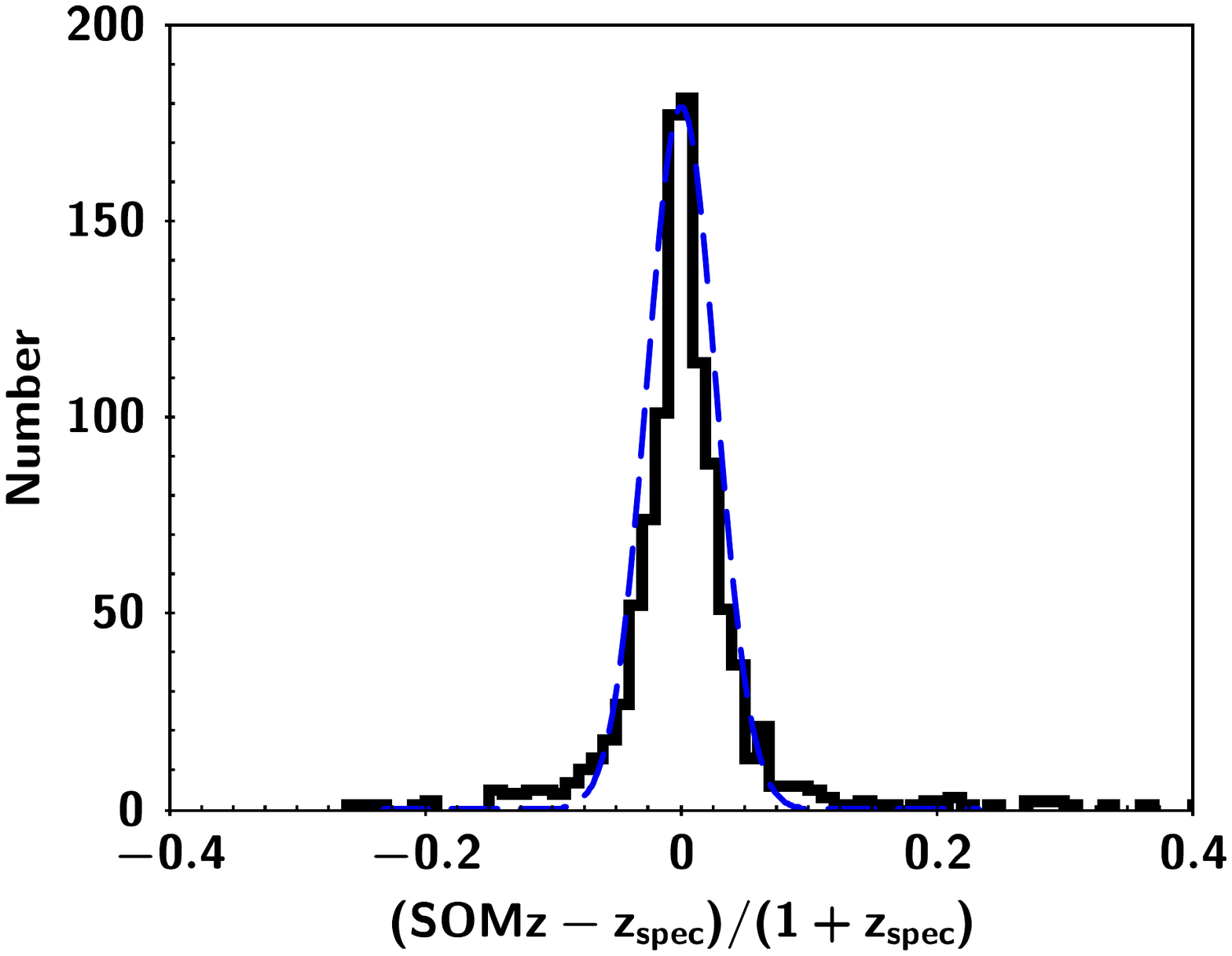} 
  \end{tabular}
  \label{figure:specz_somz}
  \caption{{\bf \emph{Top:}} Comparison of the photometric redshift predicted by the SOM with the measured Q=4 C3R2 redshifts. We define the SOM photo-z of an object to be the median of the photo-z's for all objects that best associate with its particular SOM color cell. As can be seen, the SOM photo-z estimates are mostly unbiased, with a small outlier fraction ($\sim$3.8\%, where outliers are defined as redshift errors $\geq$15\%, indicated by the thin dashed lines). {\bf \emph{Bottom:}} The distribution of $(z_{p} - z_{s})/(1 + z_{s})$ for the Q=4 sample. Overplotted is a Gaussian with $\sigma$ = 0.027, equal to the measured $\sigma_{\mathrm{NMAD}}$.}
\end{figure}

\subsection{Serendipitous Sources}

We measured the redshifts of 134 serendipitously detected sources that happened to fall in slits with primary C3R2 targets. The coordinates of these sources were identified and they were matched back to the survey catalogs. The redshifts for these sources are included in our published catalog. 

\subsection{Literature sources}
Some (unintentional) overlap with literature redshifts allows a check on our results. In COSMOS and EGS we observed 38 sources that have previously existing high-quality redshifts. Most (24) were serendipitous detections. We find an RMS discrepancy between our redshifts and the literature values of $4\times10^{-4}$. C3R2 redshifts are often higher precision than the literature values, which likely explains this small difference. There is no systematic difference between the C3R2 and literature redshifts.


\section{Redshift results and Calibration progress}


A total of 1825 sources were targeted in the 2016A observations. We identified 1131 Q=4 redshifts, 27 Q=3.5 redshifts, and 125 Q=3 redshifts. In principle, only the highest confidence redshifts would be used for calibration for cosmology; whether this restricts usable sources to those with Q=4 is worth investigating. Another 99 spectra yielded redshifts of low confidence (Q=1/2), while there were 443 failures. Of these, 409 were failure code $-91$ or $-92$, indicating that the source was too faint or lacking in identifying features, while 34 were code $-93$ or $-94$, indicating a corrupted or missing slit.

In terms of the SOM presented in M15, and using only the C3R2 sources observed in COSMOS with Q~$\geq$~3 for this analysis (911 redshifts), we have increased color space coverage by 5.4\%. Figure~\ref{figure:mag_and_z_dist} shows the \emph{i}-band magnitude distribution and redshift distribution of the 2016A ``gold" sample of Q=4 sources as well as the Q=3/3.5 sources. The distributions are very similar to the overall distribution of the unsampled cells of galaxy color space identified in M15, indicating that we are targeting the correct sources.


\subsection{SOM-based photo-z performance}
The self-organizing map colored by the median photo-z of sources per cell (the left panel of Figure~\ref{figure:colors}) effectively defines a photometric redshift estimate for each galaxy based on its position in the \emph{ugrizYJH} color space of \emph{Euclid}/\emph{WFIRST}. Figure~\ref{figure:specz_somz} compares our Q=4 spectroscopic redshifts with the redshift that would be inferred based on the SOM, with encouraging results. The normalized median absolute deviation (a dispersion measure which is not sensitive to catastrophic outliers \citep{Ilbert09,Dahlen13}) defined as \begin{equation} \sigma_{\mathrm{NMAD}} = 1.48\times \mathrm{median}\Big(\frac{\lvert z_{p}-z_{s} \rvert}{1+z_{s}}\Big) \end{equation} is 0.027 (2.7\%) for the sample, which is quite low. 

Using the standard definition of catastrophic photo-z outliers as those with $\lvert z_{p} - z_{s} \rvert/(1+z_{s}) \geq 0.15$, we measure a low outlier fraction of 3.8\%. The measured bias, defined as \begin{equation} \mathrm{mean}\Big(\frac{ z_{p}-z_{s}}{1+z_{s}}\Big) \end{equation} is $\lesssim$0.1\% after removing the catastrophic outliers. Further improvements to these results will result from folding in all spectroscopic information from C3R2 and other surveys to the \pz\ relation encoded by the SOM. Notably, these results are already competitive with or better than the photo-z results of codes tested in \citet{Dahlen13}, where the photometry used comprised 14 bands including full depth CANDELS and Spitzer data. 

\begin{figure*}[htb]
\centering
  \begin{tabular}{@{}ccc@{}}
    \includegraphics[clip, trim=1cm .5cm 1cm 0.5cm, width=.32\textwidth]{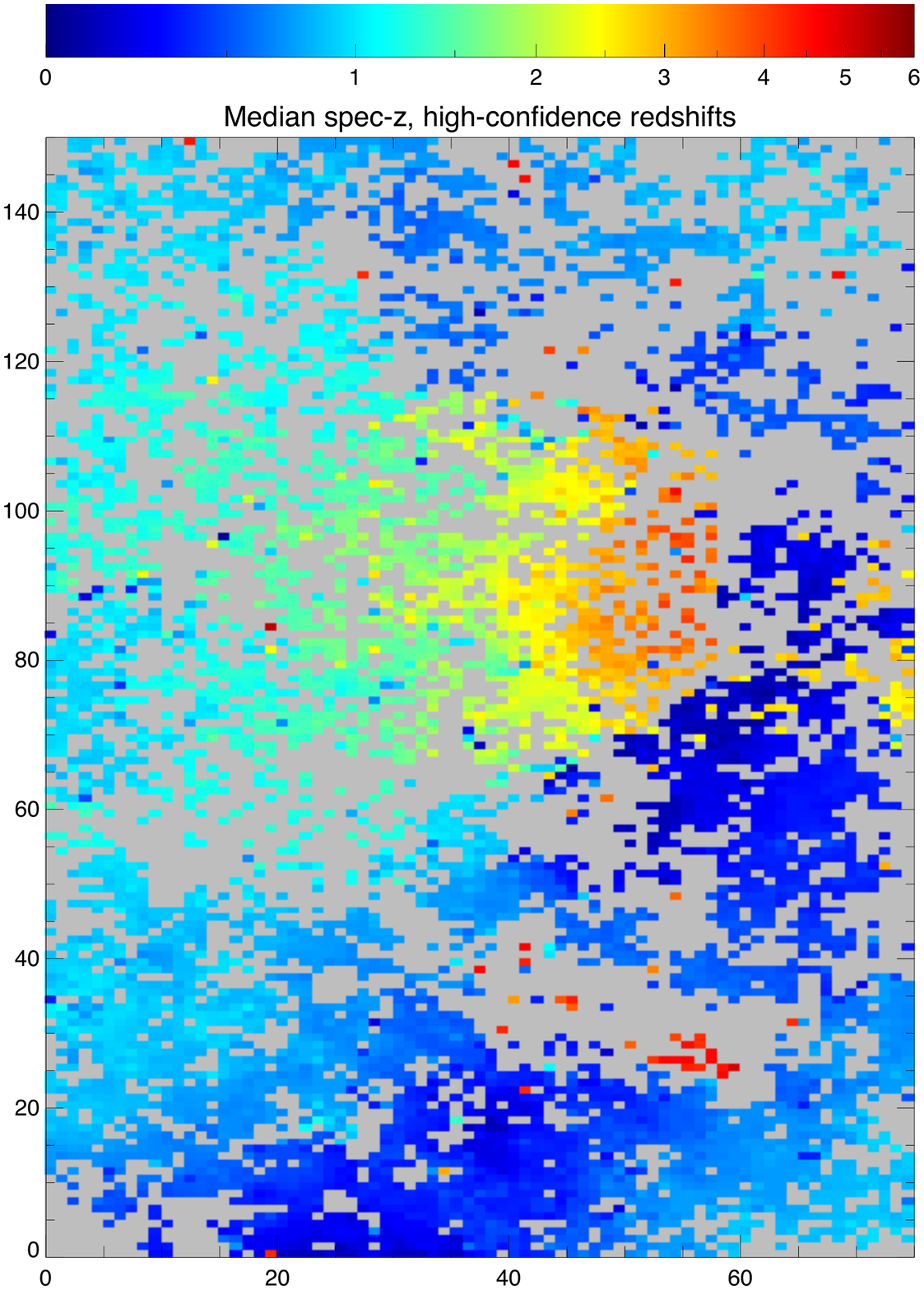} &
    \includegraphics[clip, trim=1cm .5cm 1cm 0.5cm, width=.32\textwidth]{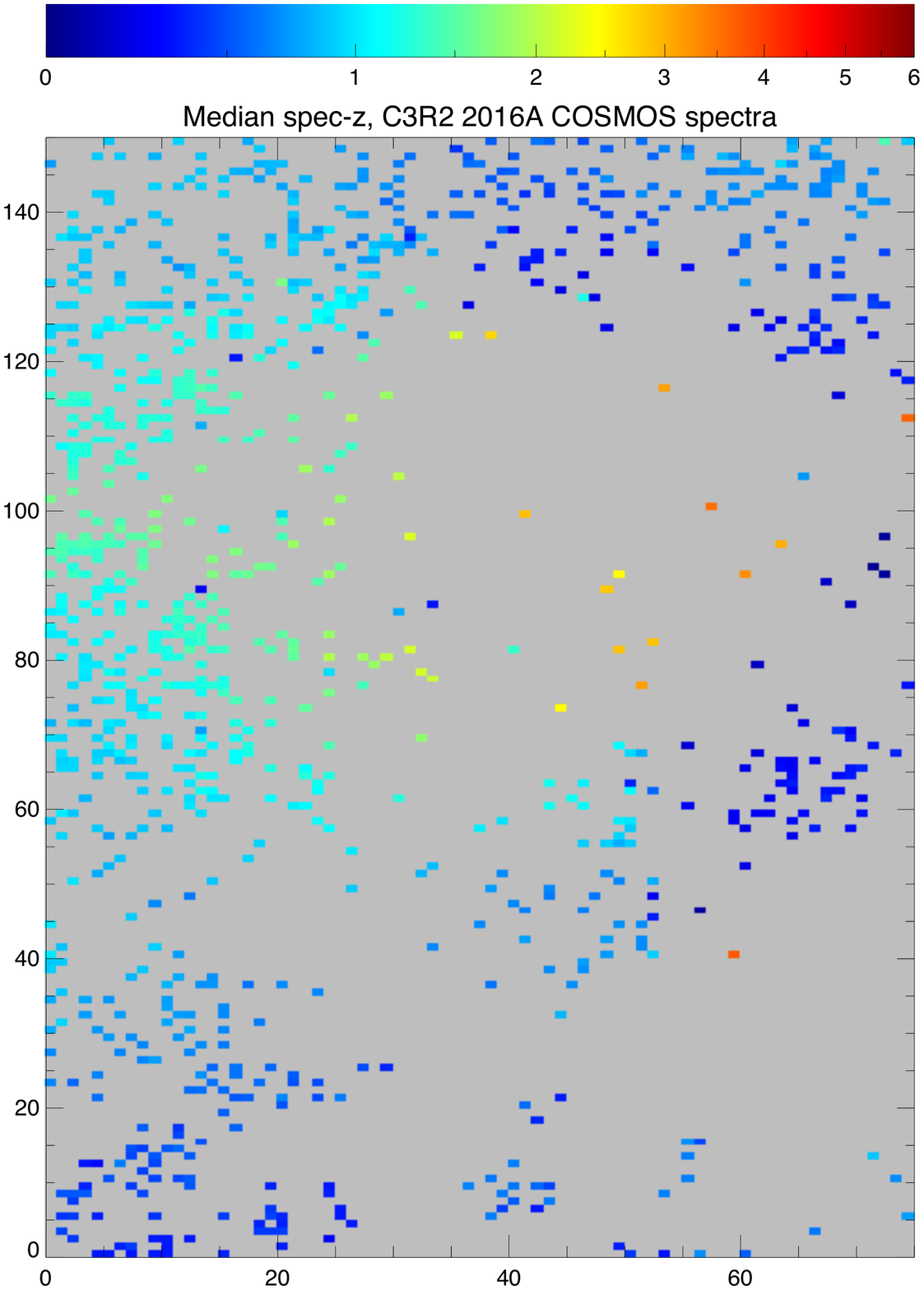} &
    \includegraphics[clip, trim=1cm .5cm 1cm 0.5cm, width=.32\textwidth]{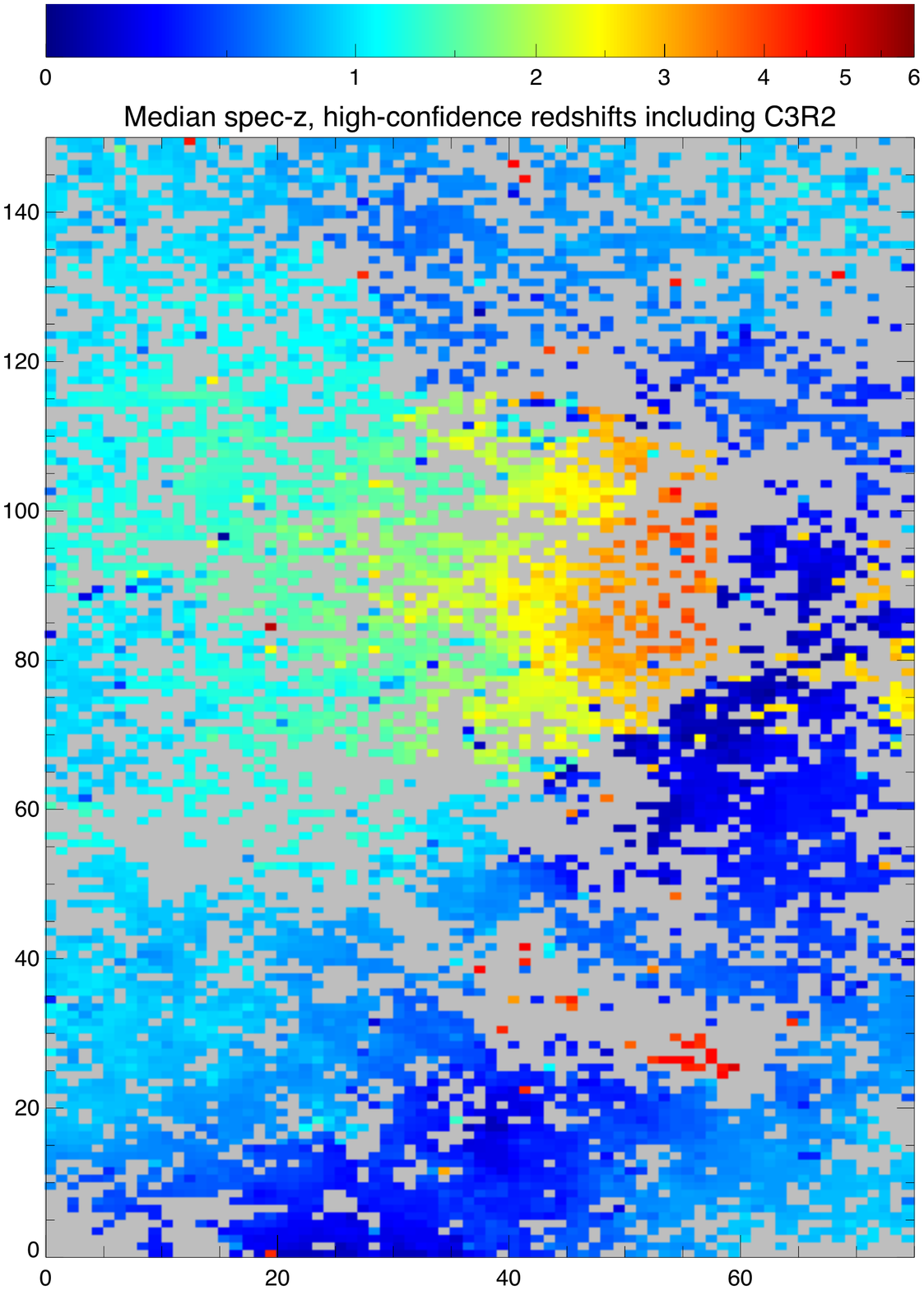} \\
  \end{tabular}
  \label{figure:som_update}
    \caption{The SOM used as the basis for the C3R2 targeting strategy in the 2016A semester, illustrating the inclusion of  the new C3R2 redshifts. {\bf \emph{Left:}} The SOM colored by the median spec-z in each color cell from preexisting spectroscopy from COSMOS. Gray cells have no preexisting high-quality redshifts. {\bf \emph{Middle:}} Distribution in color space of the 911 high-quality redshifts obtained by C3R2 in the 2016A semester in COSMOS. {\bf \emph{Right:}} The map including the high-quality redshifts added in the COSMOS field by the 2016A C3R2 nights. An additional 5.4\% of the map is measured, as well as additional constraints placed on other regions of color space. The increased density of coverage (fewer gray cells) is apparent on the right compared with the left panel. Note that this map of the relation of color to redshift is not a model, but is entirely empirical.}
\end{figure*}

While the performance we find is quite good, and may be representative of what can be achieved with a survey such as \emph{Euclid} or \emph{WFIRST}, the results depend on the depth and stability of the photometry. The photometry used to place objects on the SOM in order to estimate a photo-z in the above analysis is quite deep ($i$-band depth~$\sim$25.4 AB). The results will degrade as the photometry gets shallower or bands are lost in a manner that can be directly characterized via the SOM. A detailed study of the expected performance from the SOM-based photo-z approach will be the subject of a future paper.

\subsection{Outliers}

Out of 1079 sources with Q=4 redshifts and reliable SOM-based photo-z estimates, only 41 (3.8\%) are outliers according to the standard definition, $\lvert z_{p} - z_{s} \rvert/(1+z_{s}) \geq 0.15$. If, instead of the SOM-based photo-z, we use the photo-z for each object based on deep multiband data (e.g., the 30-band COSMOS data), we find an outlier fraction for the same sources of $\sim$3.1\%. Thus the SOM photo-z (effectively based only on the seven color \emph{Euclid}-like SEDs) performs nearly as well in terms of outlier fraction. 

We have analyzed all of the outliers on a case-by-case basis. The majority (24/41; 59\%) have \emph{individual} (rather than SOM-based) photo-z estimates more in line with the measured redshift, indicating that the color cells they belong to have real redshift scatter. For nearly all of these sources, the measured dispersion in the 30-band photo-z's within the relevant color cell is significantly larger than the median redshift dispersion per cell; in other words, these are sources that fall in more degenerate regions of the color space. The SOM can be used to identify these regions in a consistent way in order to either reject them in weak lensing analysis or direct extra spectroscopy at them to characterize the redshift distribution in those cells.

In addition, there are several other examples easily understood as Galactic stars (3) or obvious quasars/active galaxies (2), which are known not to have typical galaxy colors (total = 5/41; 12\%).  This process caught one mistaken line identification where our initial assessment of a MOSFIRE spectrum identified an isolated, narrow, strong line as [\ion{O}{3}] $\lambda 5007$ with corroborating [\ion{O}{3}] $\lambda 4959$ emission.  Subsequent analysis reveals the latter emission line to be due to poorly subtracted telluric emission, and we now identify the strong emission line as H$\alpha$ (Q=3.5).  The remaining cases seem to be genuine mismatches between the spectroscopic redshift and the photometric redshifts, both the individual photometric redshift of the galaxy and the SOM-based photometric redshift. Consideration of {\it Hubble} imaging reveals at least some of these likely due to two close-separation galaxies, where the ground-based imaging used for the photometry was unable to separate the sources.
\subsection{Increased color space coverage}
The five nights of observing in 2016A filled in $\sim$6\% of the map, in addition to existing spectroscopy which already filled $\sim$50\% (see Figure~\ref{figure:som_update}). Thus we completed $\gtrsim$10\% of the required calibration. However, some of the remaining observations may prove more challenging. Given the recent progress on bringing multiple deep fields (and their spectroscopy) onto a consistent color system, the requirements may also change to some extent, in the sense that somewhat fewer spectra are required due to the inclusion of other spectroscopic surveys. 

It should also be noted that a certain fraction of the remaining cells represent faint, red sources for which spectroscopic redshifts are prohibitively difficult to obtain with current instruments. These constitute a small ($\sim$3\%) fraction of the unsampled cells. If needed, the SOM provides a consistent method of identifying such objects and removing them from the weak lensing sample.

\scriptsize
\begin{deluxetable*}{lccccccc}
\tablecaption{Spectroscopic results.}
\tablehead{
\colhead{ID} &
\colhead{R.A.} &
\colhead{Dec.} &
\colhead{Mask} &
\colhead{Slit \#} &
\colhead{$I$ (AB)} &
\colhead{$z$} &
\colhead{Qual.}}
\startdata
UDS-3583 & 02:17:30.65 & -05:15:24.4 & UDS-m1n1 & 001 & 23.4 &  0.7877 & 4 \\
UDS-10246 & 02:17:17.55 & -05:13:06.9 & UDS-m1n1 & 002 & 23.9 &  0.8028 & 4 \\
UDS-767 & 02:17:59.05 & -05:16:21.2 & UDS-m1n1 & 003 & 25.0 &  0.5558 & 4 \\
UDS-7109 & 02:17:00.35 & -05:14:15.4 & UDS-m1n1 & 004 & 23.6 &  1.0314 & 3 \\
UDS-2276 & 02:17:52.83 & -05:15:55.2 & UDS-m1n1 & 005 & 22.9 &  0.9388 & 4 \\
UDS-8536 & 02:17:53.37 & -05:13:40.3 & UDS-m1n1 & 006 & 24.7 &  0.8619 & 4 \\
UDS-9784 & 02:17:56.80 & -05:13:15.9 & UDS-m1n1 & 009 & 23.6 &  0.8533 & 4 \\
UDS-10739 & 02:17:44.88 & -05:12:58.7 & UDS-m1n1 & 010 & 23.2 &  1.0594 & 4 \\
UDS-9730 & 02:17:32.64 & -05:13:17.4 & UDS-m1n1 & 012 & 23.8 &  1.0949 & 4 \\
UDS-12725 & 02:17:14.84 & -05:12:19.8 & UDS-m1n1 & 013 & 23.7 &  1.0351 & 4 \\
$\cdots$
\enddata
\label{table:spectroscopy}
\tablecomments{The first 10 entries of the DR1 catalog are shown. The full catalog of 1283 sources with quality flags Q$>=3$ can be found \href{http://c3r2.ipac.caltech.edu/c3r2_DR1_mrt.txt}{here}.  Note that an `s' appended to the slit number indicates a
serendipitous source.}
\end{deluxetable*}
\normalsize

\section{Conclusion}
We have presented initial results of the C3R2 survey based on five nights of Keck spectroscopy in the 2016A semester. C3R2 is designed to supplement extensive existing spectroscopy in order to provide a spectroscopic sample spanning the observed colors of galaxies to the \emph{Euclid} weak lensing photometric depth. The ultimate aim of the survey is to calibrate the color-redshift relation sufficiently to meet the requirements set by weak lensing cosmology. We estimate that the survey would require $\sim$40 Keck nights (or their equivalent) in total to meet the requirements set by \emph{Euclid}. 

In future papers we will present the updated survey strategy based on bringing multiple \emph{Euclid} calibration fields onto a consistent color system, as well as realistic tests of the performance of the method. Initial tests show that the empirical color mapping technique performs very well in reproducing $N(z)$ distributions with low bias. 


Additional data, including results from 16.5 nights allocated in 2016B as well as time allocated in 2017A and 2017B will be presented in follow-on papers. Combined with data from VLT, GTC, and MMT, we expect the calibration samples will be sufficient to meet the needs of \emph{Euclid}. Work is ongoing to understand the needs for \emph{WFIRST} calibration, but these spectra will form part of the foundation of that survey as well. Further tests and refinements of the calibration method, as well as studies to determine the optimal way to incorporate all existing spectroscopic and photometric information from deep fields into photo-z estimation using a limited set of broad band observations, are avenues of continuing research. 

\acknowledgments{The authors thank the referee, Marcin Sawicki, for a helpful report that improved this paper. D.M. would like to thank Giuseppe Longo, Audrey Galametz, Sotiria Fotopoulou, and George Helou for helpful conversations. D.M., P.C., D.S., and J.R. acknowledge support by NASA ROSES grant 12-EUCLID12-0004. J.R. and D.S. are supported by JPL, run by Caltech for NASA. F.C. acknowledges support by MINECO grant ESP2015-88861. The authors wish to recognize and acknowledge the very significant cultural role and reverence that the summit of Mauna Kea has always had within the indigenous Hawaiian community. We are most fortunate to have the opportunity to conduct observations from this mountain.}

\smallskip


\smallskip
\copyright 2017.  All rights reserved.

\bibliographystyle{apj}
\bibliography{biblio}

\clearpage
\end{document}